\documentclass[11pt, a4paper]{article}

\usepackage[utf8]{inputenc}
\usepackage{geometry}
\usepackage{hyperref}
\usepackage{booktabs} % For nicer tables
\usepackage{listings} % For code/log annexes
\usepackage{xcolor}
\usepackage{url}
\usepackage{tabularx} % Added for table width control
\usepackage{caption}  % For better caption spacing

\hypersetup{
    colorlinks=true,
    linkcolor=blue,
    filecolor=magenta,      
    urlcolor=cyan,
    citecolor=blue
}

\title{\textbf{Efficient Bitcoin Meta-Protocol Transaction and Data Discovery Through nLockTime Field Repurposing}}
\author{Nikodem Tomczak \\ nikodem.tomczak@gmail.com \\ \textit{Thulge Labs, Singapore}}
\date{December 2025}

\begin{document}

\maketitle

\begin{abstract}
We describe the Lockchain Protocol, a lightweight Bitcoin meta-protocol that enables highly efficient transaction discovery at zero marginal block space cost, and data verification without introducing any new on-chain storage mechanism. The protocol repurposes the mandatory 4-byte nLockTime field of every Bitcoin transaction as a compact metadata header. By constraining values to an unused range of past Unix timestamps greater than or equal to 500,000,000, the field can encode a protocol signal, type, variant, and sequence identifier while remaining fully valid under Bitcoin consensus and policy rules. The primary contribution of the protocol is an efficient discovery layer. Indexers can filter candidate transactions by examining a fixed-size header field, independent of transaction payload size, and only then selectively inspect heavier data such as OP\_RETURN outputs or witness fields. The Lockchain Protocol applies established protocol design patterns to an under-optimised problem domain, namely transaction discovery at scale, and does not claim new cryptographic primitives or storage methods.
\end{abstract}

\section{Introduction}
For current Bitcoin meta-protocols to function efficiently, robust discovery and indexing mechanisms are required to identify which transactions belong to which protocol and which outputs represent protocol-relevant assets. In most deployed systems, discovery requires scanning and parsing protocol-specific payloads embedded in script, OP\_RETURN outputs, or witness data. This process scales linearly with the amount of transaction data that must be inspected.

Counterparty, Omni Layer, and Colored Coins rely on scanning OP\_RETURN outputs for protocol-specific formats and magic bytes at the beginning of the embedded data \cite{bartoletti2017}. Ordinals Inscriptions require parsing SegWit transactions and their witness data to detect a specific envelope pattern, often nested inside Taproot scripts \cite{rodarmor2023}. BRC-20 protocols further require parsing JSON-formatted payloads contained within these envelopes \cite{domo2023}. In all cases, indexers must ingest and inspect the full relevant payload to determine whether a transaction is protocol-relevant.

Full indexer synchronisation commonly takes tens of hours, requires tens of gigabytes of auxiliary database storage, and incurs non-trivial infrastructure costs. Because of this, most users depend on hosted indexers, increasing centralisation risks. The core inefficiency is the absence of a low-cost prefiltering mechanism that allows indexers to discard the vast majority of transactions without parsing large payloads. This inefficiency also makes searching for transactions with specific protocol-level metadata cumbersome, as it requires repeated deep inspection of transaction data.

The Lockchain Protocol does not aim to replace existing storage mechanisms such as OP\_RETURN or witness-based data embedding. Instead, it introduces a unifying metadata layer that improves the discoverability and indexing efficiency of such protocols. The approach leverages unused semantic space in the nLockTime field by selecting values in a past timestamp range that are valid for immediate inclusion in blocks. By doing so, the protocol enables a filtering step that depends only on a fixed-size header field. This results in a substantial reduction in the amount of transaction data that must be parsed during discovery, while remaining fully compatible with existing consensus and policy rules. We build on field repurposing precedents and use established patterns to solve practical optimisation problems. The protocol is intentionally conservative in scope and focuses on architectural efficiency rather than introducing new cryptographic constructions.

\section{The nLockTime Field: Function and Repurposing}
The nLockTime field is a 32-bit unsigned integer defined in the original Bitcoin transaction format and present in every transaction \cite{nakamoto2008}. Its purpose is to restrict the earliest time or block height at which a transaction may be included in a valid block. Values below 500,000,000 are interpreted as block heights, while values equal to or above this threshold are interpreted as Unix timestamps, as defined in Bitcoin Core consensus logic.

Enforcement of nLockTime depends on the nSequence field of transaction inputs. For the nLockTime constraint to be enforced, at least one input must have an nSequence value below 0xFFFFFFFF. If all inputs have nSequence set to 0xFFFFFFFF, the nLockTime value is ignored by consensus. In practice, many wallets deliberately set non-zero nLockTime values even when no strict time locking is required. One common motivation is fee-sniping mitigation, where wallets set nLockTime to a recent block height to discourage miners from attempting short-range chain reorganisations for transaction fee capture \cite{bip125}.

Modern wallet behaviour also commonly enables Replace-By-Fee to avoid transactions becoming stuck in the mempool. BIP 125 specifies that signalling opt-in Replace-By-Fee requires at least one input to have nSequence below 0xFFFFFFFE, which in turn activates nLockTime enforcement \cite{bip125}. As a result, transactions that support Replace-By-Fee must use an nLockTime value that is already satisfied at the time of broadcast if immediate settlement is desired.

Beyond wallet-level policy, nLockTime and related timelock primitives are fundamental components of Bitcoin smart contracts. Transaction-level and output-level timelocks are used in constructions such as payment channels and Hash Time-Locked Contracts, most notably in the Lightning Network \cite{poon2016}. These uses rely on nLockTime’s consensus semantics rather than wallet policy behaviour.

This interaction between fee-sniping mitigation, Replace-By-Fee signalling, and contract-level enforcement creates a practical constraint. To support Replace-By-Fee while remaining immediately mineable, nLockTime must be set to a value in the past. From the perspective of Bitcoin consensus, such a value is valid and imposes no effective time lock. The Lockchain Protocol exploits this requirement by encoding metadata into nLockTime values that represent timestamps long before the existence of Bitcoin, where the time-locking semantics are effectively meaningless but consensus validity is preserved.

\section{nLockTime Field Structure and Encoding}
The Lockchain Protocol divides the 4-byte nLockTime field into a structured metadata header consisting of four single-byte fields. Byte 1 is the Magic or Signal byte, Byte 2 indicates the Type, Byte 3 indicates the Variant or Method, and Byte 4 is the Sequence number. This structure provides over four billion theoretical combinations, though we constrain this to a pragmatic past timestamp range to ensure consensus validity and semantic correctness.

While the 1-1-1-1 byte division provides balanced flexibility and readability, protocol designers can adopt alternative schemes optimized for specific use cases. A 3-1 byte split, allocating 24 bits for identifier and 8 bits for sequence, supports over 16 million unique protocol and type combinations with up to 256 sequential transactions per protocol instance. A 2-2 byte split, with 16 bits for protocol identification and 16 bits for sequence, supports 65,536 distinct protocols with up to 65,536 sequential transactions each, suitable for protocols requiring extensive sequencing. A 2-1-1 split could allocate 16 bits for magic and type combined, 8 bits for variant, and 8 bits for sequence, serving protocols with complex type hierarchies but simpler sequencing needs.

At the extreme, bit-level encoding is theoretically possible within the pragmatic timestamp range. Individual bits across all four bytes could encode a 32-bit state machine, capability flags, or nested hierarchies. Bit-level encoding requires bitwise operations such as shifts, masks, and AND/OR operations rather than simple byte extraction, increasing indexer implementation complexity slightly. The performance impact is negligible on modern hardware, as bitwise operations are single-cycle instructions, but code readability and maintainability are reduced. Byte-aligned fields such as 1-1-1-1, 2-2, or 3-1 offer a pragmatic balance of flexibility, performance, and human readability. The key constraint for any encoding scheme is that the final 32-bit value must fall within the pragmatic past timestamp range from 500,000,000 to the current Unix timestamp.

The protocol identifies a pragmatic range of 4-byte nLockTime values based on Bitcoin consensus rules. The value must be greater than 500,000,000 to be interpreted as a timestamp and less than the current time to ensure immediate settlement and compatibility with Replace-By-Fee. The minimum timestamp value is 500,000,000 (hex 0x1DCD6500, Jan 1985). The current timestamp as of late 2025 is approximately 1,763,000,000 (hex 0x6918C260). This establishes the working range for the Magic Signal. To ensure the 32-bit integer is always interpreted as a valid past timestamp regardless of subsequent bytes, the first byte must fall within a specific range. The lowest safe byte is 0x1F, as 0x1F000000 equals 520,093,696 decimal, safely above the 500,000,000 threshold. Using 0x1E would result in approximately 469 million, which could be interpreted as a block height in the future, effectively locking the transaction. The highest safe byte is 0x68, as 0x68FFFFFF equals 1,761,607,679 decimal, safely below the current 2025 timestamp threshold. This time-dependent nature of the highest safe byte should be acknowledged as an edge condition: as time progresses, the available signalling space technically grows, not shrinks, and the protocol remains valid.

The usable Magic Signal range 0x1F to 0x68 also maps nearly perfectly to printable ASCII characters, with 0x20 through 0x68 covering space, numbers, uppercase letters, and lowercase a through h. This allows limited human-readability in hex dumps, though consensus correctness is primary. The remaining bytes provide protocol flexibility. The Type byte allows for 256 distinct protocol types, the Variant byte provides 256 sub-categorizations, and the Sequence byte enables ordering for up to 256 fragments, or can be combined with the Variant byte for a 16-bit sequence allowing for 65,536 fragments.

\section{Repurposing Bitcoin’s Transaction Fields}
The concept of repurposing protocol fields for metadata storage has precedents in Bitcoin history. Fields that were underutilized for their original purpose or where new uses would not conflict with existing functionality have been coopted to enhance the protocol. For example, the nSequence field, originally intended for transaction replacement signalling, was later repurposed. BIP 68 (Relative lock-time using consensus-enforced sequence numbers) introduced relative lock-time to the sequence number field, enabling a signed transaction input to remain invalid for a defined period after confirmation of its corresponding outpoint \cite{bip68}. This preserved the sequence number’s use while adding new functionality. In addition, the BIP authors clearly thought the remaining unused bits in nSequence could be repurposed further for new use cases. Similarly, BIP 125 (Opt-in Replace-By-Fee) allowed the nSequence field to signal RBF eligibility, indicating that a transaction could be replaced if any input has an nSequence number below 0xFFFFFFFE \cite{bip125}.

The nVersion field was repurposed for soft-fork signalling in BIP 9 (Version bits with timeout and delay, 2015) and BIP 8 (Version bits with lock-in by height, 2017) \cite{bip9}. The nLockTime field itself has seen multiple interpretation changes, although it has not previously been used for meta-protocol discovery. The Lockchain Protocol proposes using nLockTime as a discovery beacon while maintaining consensus validity.

\begin{table}[ht]
\centering
\caption{Repurposing of Transaction Fields}
\label{tab:repurposing}
\begin{tabularx}{\textwidth}{@{}l l X l@{}}
\toprule
\textbf{Field} & \textbf{Original Use} & \textbf{Repurposed For} & \textbf{Year} \\
\midrule
nSequence & Tx replacement (unused) & Relative timelocks & 2015 (BIP 68) \\
nSequence & Tx replacement (unused) & Opt-in RBF Signaling & 2015 (BIP 125) \\
nVersion & Protocol version & Soft fork signaling & 2015 (BIP 9) \\
nLockTime & Time-locking & Meta-protocol discovery & 2025 (Proposed) \\
\bottomrule
\end{tabularx}
\end{table}

There are several reasons why nLockTime was not historically used for meta-protocol discovery. First, it has a well-defined purpose. it enforces time constraints on transaction validity. Arbitrary past timestamps are semantically meaningless from a consensus perspective. Second, since 2014, OP\_RETURN has been the standard for on-chain metadata storage \cite{bartoletti2017}. Early OP\_RETURN was limited to 80 bytes per transaction, while modern node policies allow up to 100,000 bytes. Witness data has also become a popular, discounted storage medium. It was only when the number of meta-protocols increased significantly that indexing inefficiencies became apparent, motivating exploration of alternative discovery mechanisms.

There has been little comprehensive analysis of nLockTime usage patterns. b10c \cite{b10c2020} observed the “stair-pattern” in time-locked transactions, reflecting wallet behaviour where nLockTime is set near the current block height. Approximately 80 percent of transactions have nLockTime equal to 0, meaning no locktime. Of the remainder, most are set to current or near-current block heights, with occasional future timestamps for payment channels and escrows. The presence of nLockTime values set far in the past is minimal, making it suitable for use as a metadata beacon with low collision probability.

Other relevant considerations include CoinJoin implementations such as Wasabi and JoinMarket, which manipulate nLockTime for coordination purposes, and miners’ anti-spam heuristics that occasionally inspect unusual locktime values. While these behaviours are rare, they indicate that nLockTime manipulation has precedent and should be considered when evaluating potential collisions or noise.

Finally, the protocol assumes SegWit transactions, since modern meta-protocols operate primarily within this context. This ensures compatibility with contemporary transaction structures, witness data, and dust-output handling.

\section{Lockchain Protocol Proof-of-Concepts}
The Lockchain Protocol proofs-of-concept demonstrate the practical application of the 4-byte nLockTime metadata header for verifiable on-chain data discovery. A Simple Sharding Model (POC 1) uses nLockTime for sequencing OP\_RETURN data. The nLockLighthouse models (POC 2, 2.5, and 3) combine the nLockTime beacon with a genesis transaction and spendable dust tokens to represent fractional, transferable ownership. The primary innovation of the Lockchain Protocol is a highly efficient off-chain discovery layer, allowing indexers to find assets by scanning only the 4-byte nLockTime field. This enables O(1) complexity relative to the transaction data size, in contrast to terabytes of witness data or OP\_RETURN scripts requiring deep inspection in conventional protocols. Additionally, linking a voluntary registry of Magic/Type allocations reduces noise and potential collisions during discovery.

\subsection{POC 1: Simple Sharding with nLockTime Sequencing}
The Simple Sharding proof-of-concept (POC 1) demonstrates the most basic application of the nLockTime field for transaction sequencing and protocol discovery. This model is designed to show how the 4-byte nLockTime header can be used to create an ordered sequence of transactions, each carrying a fragment of data in its OP\_RETURN output. The primary purpose is to validate the core mechanism, specifically using nLockTime as a discovery beacon and sequence identifier, rather than to provide a practical data storage solution. The protocol operates by first calculating how many transactions are required to store the input data.

Given an OP\_RETURN limit, which is 80 bytes in the standard case though configurable up to 100KB in some node policies, and a hash size typically 32 bytes for SHA256, the usable payload per transaction is the OP\_RETURN limit minus the hash size. While modern policies may allow up to 100KB, enabling single-transaction storage for many files, this specific experiment intentionally utilizes a smaller standard limit to demonstrate the protocol's sequencing robustness over a larger number of transactions. As detailed in Annex 1, a 1,120-byte file was processed with a 48-byte effective payload per shard, requiring 24 transactions or shards to complete the sequence. The implementation includes a timestamp proof by hashing the concatenation of the file data and the current Unix timestamp, creating a verifiable proof of when the data existed in this specific form.

To ensure sufficient UTXOs are available for all shard transactions, the protocol begins with a funding transaction. This funding transaction serves dual purposes: it embeds the first data shard (shard 0) in its OP\_RETURN output along with the full data hash, and it creates N-1 dust outputs (one for each remaining shard) that will be consumed as inputs by the subsequent shard transactions. The funding transaction's nLockTime structure is [Magic: 0x4C][Type: 0x01][Sequence: 0x0000], where the magic byte 0x4C ('L') serves as the protocol identifier (an Integer from 31 to 104), the type byte 0x01 indicates this is POC 1, and the sequence number 0x0000 indicates this is shard 0. Each subsequent shard transaction consumes one of the funding outputs as input, includes an OP\_RETURN output containing both the full data hash and the shard's data fragment, and sets its nLockTime to increment the sequence number. Shard 1 has nLockTime [0x4C][0x01][0x00][0x01], shard 2 has [0x4C][0x01][0x00][0x02], and so forth up to shard N with sequence number 0x0017 in the 1,120-byte example. This creates a verifiable sequence where an indexer can discover all shards by scanning for the magic byte 0x4C and type byte 0x01, then reassemble the data in the correct order using the sequence numbers.

The cost structure is linear with the number of shards. As shown in the experimental data, for 24 shards at approximately 1,200 satoshis per transaction, the total network fee was 28,800 satoshis. This makes POC 1 using small shards expensive for data storage and impractical for files larger than a few kilobytes unless the 100KB policy is leveraged. The OP\_RETURN outputs in POC 1 are, by Bitcoin consensus, provably unspendable. The only spendable outputs are the change outputs returned to the issuer after each transaction. This means POC 1 does not create transferable assets or ownership tokens, as it is purely a data commitment and sequencing demonstration. There is no mechanism to prevent others from copying the data and creating their own shard sequence, and there is no way to transfer ownership of the data commitment to another party. The protocol also does not include any protection against spending the dust outputs prematurely, which would break the shard sequence if done before all shards are broadcast. POC 1 just demonstrates the most basic nLockTime discovery mechanism and sequence numbering capability but makes no attempt to solve the security, ownership, or cost-efficiency problems inherent in on-chain data storage. It serves as a foundation to show that nLockTime can be reliably used for transaction ordering and that indexers can efficiently discover protocol transactions by scanning the 4-byte header rather than parsing full transaction data.

\subsection{POC 2: Single-Transaction Static Asset with Merkle Root}
The Single-Transaction Static Asset (POC 2) addresses the primary cost inefficiency of POC 1 by eliminating on-chain data storage entirely and replacing it with a cryptographic proof of data integrity. Instead of storing data fragments in OP\_RETURN outputs across multiple transactions, POC 2 stores only a 32-byte Merkle root in a single transaction's OP\_RETURN output. This Merkle root serves as a compact, verifiable fingerprint of the entire file, allowing anyone with the original data to independently reconstruct the Merkle tree and confirm that it matches the on-chain commitment \cite{rosenfeld2012}.

The protocol begins by dividing the input file into N chunks, where N is specified by the user rather than being derived from data size. This design decision decouples the cost from file size, meaning a 184KB file divided into 10 chunks costs the same as a 4MB file divided into 10 chunks, because both result in a single transaction with 10 dust outputs. Each chunk is hashed to create a leaf node of the Merkle tree. Pairs of these leaf hashes are concatenated and hashed recursively, proceeding up the tree until a single hash remains: the Merkle root. This root is embedded in the genesis transaction's OP\_RETURN output (output 0). For each chunk, the genesis transaction creates a dust output (546 satoshis by default, though configurable) sent to the issuance address. These outputs, numbered as outputs 1 through N in the genesis transaction, represent fractional ownership of the asset. Unlike POC 1's unspendable OP\_RETURN data, these dust outputs are standard P2WPKH (Pay-to-Witness-Public-Key-Hash) UTXOs that can be spent and transferred like any other Bitcoin output. A protocol fee output is also included, calculated as 10 percent of the total dust outputs with a minimum of 546 satoshis to avoid creating dust that would be rejected by network relay rules.

The nLockTime structure for POC 2 is [Magic: 0x4C][Type: 0x02][Variant: 0x73][Sequence: 0x01], where the variant byte 0x73 ('s') indicates a single-transaction model, distinguishing it from the multi-transaction POC 2.5 variant. The magic and type bytes allow indexers to discover POC 2 assets by scanning nLockTime values, while the Merkle root in the OP\_RETURN output provides the content commitment. The sequence byte can be used to version the protocol or to number multiple asset issuances by the same issuer. The cost structure demonstrates POC 2's efficiency advantage. As documented in Annex 2, for a 10-chunk asset, the total cost is approximately 1,850 satoshis, plus the value of the dust tokens themselves. This cost remains constant regardless of file size. This is orders of magnitude cheaper than storing the same file using POC 1's OP\_RETURN method.

However, POC 2 introduces a critical limitation: there is no cryptographic binding between the issuer's identity and the content. The genesis transaction ID, which represents the issuer, and the Merkle root, which represents the content, exist as independent pieces of information. An indexer can verify that a Merkle root was committed to in a specific transaction, but there is no cryptographic proof linking the two. This means an attacker could, in theory, take someone else's Merkle root and create their own genesis transaction claiming it as their own issuance, or conversely, take someone else's genesis transaction ID and claim it commits to different content. While indexers can apply first-seen rules or trust the blockchain's temporal ordering, there is no mathematical proof preventing such substitution attacks. The dust outputs in POC 2 are also vulnerable to accidental spending. Unlike protocols that use covenant-style restrictions or multisignature schemes, POC 2's dust outputs are standard P2WPKH outputs controlled solely by the issuer's private key. If the issuer's wallet inadvertently consolidates UTXOs or spends one of these dust outputs in an unrelated transaction, the fractional ownership structure is broken. There is no on-chain mechanism to warn the wallet that certain UTXOs represent protocol assets rather than ordinary bitcoin. This limitation is shared with other dust-token-based protocols like early Counterparty implementations. POC 2 demonstrates that nLockTime-based discovery can be combined with Merkle tree commitments to create cost-efficient asset representations on Bitcoin. The single-transaction model is simpler to implement and broadcast than POC 1's multi-transaction sequence, and the cost savings are substantial. However, the lack of issuer-content binding and the absence of spending protection expose security gaps that subsequent POCs address. POC 2 establishes the basic asset issuance pattern, a Merkle root commitment plus dust outputs, that serves as the foundation for the more secure two-transaction and multisignature models that follow.

\subsection{POC 2.5: Two-Transaction Static Asset with Cryptographic Binding}
The Two-Transaction Static Asset with Cryptographic Binding (POC 2.5) extends the Static Asset Lighthouse model (POC 2) by introducing a robust issuer-to-content binding mechanism across two separate transactions. This proof-of-concept addresses a fundamental security consideration: in POC 2, while the Merkle root proves the integrity of the data, there is no cryptographic link between the issuer's identity (represented by the genesis transaction ID) and the content itself. POC 2.5 solves this by creating a binding hash that ties the issuer to the content, preventing content substitution attacks and enabling verifiable provenance chains.

The protocol operates through two distinct transactions with sequential nLockTime values. The first transaction (Genesis Transaction, TX1) establishes the issuer's commitment to the content by embedding the Merkle root in an OP\_RETURN output. The nLockTime structure for TX1 is [Magic: 0x4C][Type: 0x03][Variant: 0x74][Sequence: 0x00], where the variant byte 0x74 ('t') indicates a two-transaction model. This transaction serves as the anchor point and contains all protocol fees for both transactions combined. The protocol fee calculation is based on 10 percent of the default network fee (covering TX1) plus 10 percent of the total dust outputs (covering TX2), with a minimum threshold of 546 satoshis to avoid creating dust outputs that would be rejected by the network.

After TX1 is broadcast and propagates through the network, the protocol constructs a binding hash using the formula: $binding\_hash = SHA256(genesis\_txid + merkle\_root)$. This binding hash cryptographically links the issuer's identity (the unique genesis transaction ID, which cannot be forged or predicted) to the specific content (the Merkle root). The binding hash serves as proof that the tokens created in TX2 are authentically tied to both the issuer who created TX1 and the exact content committed to in that transaction. An attacker cannot substitute different content or claim to be a different issuer without invalidating this binding.

The second transaction (Tokenization Transaction, TX2) creates the spendable ownership tokens and embeds the binding hash in its OP\_RETURN output. The nLockTime structure for TX2 is [Magic: 0x4C][Type: 0x03][Variant: 0x74][Sequence: 0x01], where the incremented sequence number 0x01 indicates this is the second transaction in the protocol sequence. TX2 contains no protocol fee output, as all protocol fees were collected in TX1. For each chunk of the original file, TX2 creates a 546-satoshi (or higher, depending on network policy) dust output sent to the issuance address. These outputs represent fractional, transferable ownership of the asset. The binding hash in TX2's OP\_RETURN provides indexers with the cryptographic proof they need to verify the complete chain: TX2 tokens to binding hash to TX1 genesis to Merkle root to original content.

The sequential nLockTime values (TX1 sequence 0x00, TX2 sequence 0x01) serve multiple purposes. First, they enable indexers to efficiently discover both transactions belonging to the same asset issuance by scanning for matching magic bytes, type bytes, and variant bytes. Second, they establish an unambiguous temporal ordering, ensuring that indexers can reconstruct the protocol flow correctly. Third, they demonstrate the protocol's capability for multi-transaction sequences, which could be extended to more complex use cases requiring three or more linked transactions. The cryptographic binding mechanism provides several security guarantees. First, it prevents an attacker from taking someone else's genesis transaction and claiming their own tokenization transaction belongs to it, as the binding hash would not match. Second, it prevents content substitution, as an attacker cannot change the Merkle root in TX1 without invalidating the binding hash in TX2. Third, it creates a verifiable chain of custody that indexers can validate without trusting any intermediary. Any party with access to the blockchain can independently verify that TX2's binding hash was correctly derived from TX1's genesis transaction ID and Merkle root.

The cost structure for POC 2.5 differs from POC 2 due to the two-transaction model. As shown in Annex 3, for a 10-chunk asset, the total cost is approximately 5,500 satoshis. Despite the additional transaction, this cost remains independent of file size, demonstrating the same efficiency advantage over data-storage-based protocols. The two-transaction model introduces a dependency: TX2 cannot be broadcast until TX1 has been confirmed (or at minimum, accepted into the mempool and propagated). The implementation includes a waiting period after TX1 broadcast to allow network propagation before constructing TX2. Additionally, the wallet must be refreshed between transactions to ensure the change output from TX1 is available as an input for TX2. These requirements add operational complexity compared to the single-transaction POC 2 but provide the security benefit of cryptographic binding. POC 2.5 demonstrates that the nLockTime beacon mechanism can coordinate multi-transaction protocols while maintaining the zero-cost discovery advantage. Indexers can find both transactions by scanning for the protocol's magic byte and type byte, then use the sequence numbers to correctly order them. The binding hash provides a verifiable link that allows indexers to confidently associate TX2's ownership tokens with TX1's content commitment, creating a complete, auditable provenance chain.

\subsection{POC 3: Multisig-Protected Static Assets}
The "Multisig-Protected Static Asset" proof-of-concept (POC 3) extends the two-transaction model from POC 2.5 by adding a critical safety mechanism based on a multisignature protection of ownership tokens. While POC 2.5 successfully created a cryptographically bound link between issuer identity and content, it left tokens vulnerable to accidental spending by wallets that don't understand the protocol. A user consolidating UTXOs or sweeping their wallet could inadvertently destroy fractional ownership tokens, mistaking them for ordinary 546-satoshi dust outputs. POC 3 solves this by making each token a 2-of-2 multisignature output requiring both the owner's signature and a service co-signature, preventing standard Bitcoin wallets from spending these UTXOs even if they try.

The protocol introduces a variant-based transaction type system encoded in the nLockTime field's third byte. Unlike POC 2.5's fixed variant byte, POC 3 uses three distinct variants to identify different transaction types within the protocol: variant 'g' (0x67) for the Genesis transaction that commits to the Merkle root, variant 't' (0x74) for the Tokenization transaction that creates the multisig-protected tokens, and variant 'x' (0x78) for Transfer transactions that move token ownership. This variant system allows indexers to immediately identify a transaction's role within the protocol by examining the nLockTime field alone, without needing to parse the transaction's inputs and outputs. The fourth byte (sequence) serves different purposes depending on the variant: for genesis and tokenization transactions it acts as a basic sequence number (0x00 and 0x01 respectively), while for transfer transactions it increments with each transfer, creating an on-chain transfer count that provides a verifiable ownership history.

The Genesis transaction (TX1) operates identically to POC 2.5's genesis transaction, embedding the 32-byte Merkle root in an OP\_RETURN output and collecting all protocol fees for the entire asset creation sequence. The nLockTime structure is [Magic: 0x4C][Type: 0x03][Variant: 0x67][Sequence: 0x00], where the variant byte 0x67 ('g') explicitly marks this as a genesis commitment. After TX1 propagates through the network and receives a transaction ID (genesis\_txid), the protocol computes the binding hash using the same formula as POC 2.5: $binding\_hash = SHA256(genesis\_txid + merkle\_root)$. This binding hash cryptographically links the issuer's identity to the specific content, creating a verifiable provenance chain. The Tokenization transaction (TX2) represents POC 3's primary innovation. Instead of creating standard P2WPKH dust outputs as in POC 2 and POC 2.5, TX2 creates P2WSH (Pay-to-Witness-Script-Hash) outputs locked by 2-of-2 multisignature scripts. Each token's redeem script follows the pattern: OP\_2 Owner\_Pubkey Service\_Pubkey OP\_2 OP\_CHECKMULTISIG. The script requires two valid signatures—one from the current owner and one from a designated service—before the UTXO can be spent. The nLockTime structure for TX2 is [Magic: 0x4C][Type: 0x03][Variant: 0x74][Sequence: 0x01], where variant 't' (0x74) identifies this as the tokenization transaction.

The multisignature protection mechanism provides two critical guarantees. First, it prevents accidental spending by standard wallets. A wallet that attempts to spend a token UTXO will fail because it only controls the owner's private key, not the service's private key. The wallet cannot generate the second required signature, so the transaction will be invalid and rejected by the network. This protects users from inadvertently destroying their asset ownership when performing routine wallet operations. Second, it enables transfer validation by the service. When an owner wants to transfer a token to a new recipient, they construct a partially signed transaction (PSBT) with their signature and submit it to the service. The service can then validate the transfer by checking that the nLockTime variant is 'x' (0x78) and that the sequence number has incremented by exactly one from the previous transfer. It validates that the new output is a proper 2-of-2 multisig with the service and the recipient. Only after all validations pass does the service add its signature and broadcast the fully signed transaction.

Transfer transactions introduce the third variant type and demonstrate how the sequence byte tracks ownership history. A transfer transaction has the nLockTime structure [Magic: 0x4C][Type: 0x03][Variant: 0x78][Sequence: TT], where variant 'x' (0x78) identifies this as a transfer, and the sequence byte TT represents the transfer count. The first transfer has sequence 0x01, the second transfer has sequence 0x02, and so forth, creating an immutable on-chain record of how many times the token has changed hands. The cost structure for POC 3 matches POC 2.5's base costs with negligible overhead from multisig scripts. As shown in Annex 4, the total creation cost for a 10-chunk asset is approximately 2,869 satoshis (TX1 fee + TX2 fee + Protocol fee). Transfer costs are similarly modest. The critical efficiency advantage remains unchanged from POC 2.5: costs scale with the number of tokens (chunks) rather than file size, making POC 3 economically viable for large files that would be prohibitively expensive under data-storage-based protocols. POC 3 demonstrates that ownership tokens can be protected from accidental spending while maintaining the UTXO-based ownership model and cost efficiency of POC 2.5.

\section{Comparative Efficiency and Discovery Analysis}
The Lockchain Protocol is compared against existing Bitcoin meta-protocols with respect to discovery efficiency, indexer requirements, data integrity, asset ownership, on-chain footprint, and cost model. The primary differentiator is the discovery mechanism. While conventional protocols require linear scanning of the blockchain data (witness or script), Lockchain allows for constant-time O(1) filtering based on the nLockTime field in the transaction header.

For protocols like Ordinals or BRC-20, discovering an asset requires deep inspection of every transaction’s witness data. If a transaction is 4 MB, the indexer must read all 4 MB, and as the blockchain grows, this cost scales linearly (O(N)). Lockchain, by contrast, examines only the fixed 4-byte nLockTime field. The indexer performs a single integer comparison (e.g., $nLockTime \geq 500,000,000$) per transaction, regardless of the transaction size. This enables immediate filtering of non-Lockchain transactions, ignoring most of the blockchain’s data volume during discovery.

To further reduce noise and potential collisions, a voluntary registry of Magic/Type allocations can be maintained in a public repository (e.g., GitHub, DNS-style naming, or SLIP-44-like numeric assignments). By consulting this registry, indexers can prioritize known protocol identifiers and avoid accidental collisions from otherwise valid transactions.

\begin{table}[ht]
\centering
\caption{Discovery and Efficiency Comparison}
\label{tab:comparison}
\footnotesize
\begin{tabularx}{\textwidth}{@{}X X X X X@{}}
\toprule
\textbf{Feature} & \textbf{Lockchain (Lighthouse)} & \textbf{Ordinals} & \textbf{BRC-20} & \textbf{Counterparty / Omni} \\
\midrule
Discovery Mechanism & O(1) Header Scan (nLockTime) & O(N) Deep Witness Scan & O(N) Deep Witness Scan & O(N) Output Scan (OP\_RETURN) \\
\addlinespace
Indexer Requirement & Headers + Tx Metadata Only & Full Chain + Witness Data & Full Chain + Witness Data & Full Chain + Output Scripts \\
\addlinespace
Data Integrity & Merkle Root (Hash) & Full On-Chain Data & Text/JSON in Witness & Data in OP\_RETURN \\
\addlinespace
Asset Ownership & Spendable Dust UTXO & Spendable Dust UTXO & Spendable Dust UTXO & Virtual Ledger Balance \\
\addlinespace
On-Chain Footprint & Minimal (Header + 32B Hash) & Heavy (Full File Size) & Heavy (JSON Text) & Moderate (80B OP\_RETURN) \\
\addlinespace
Cost Model & Fixed (Independent of file size) & Linear (Scales w/ size) & Linear (Scales w/ size) & Linear (Scales w/ size) \\
\bottomrule
\end{tabularx}
\end{table}

Lockchain’s efficiency is achieved by leveraging SegWit transactions (assumed implicitly throughout the proofs-of-concept), which ensures witness data is separate and discounted, but discovery only depends on the header. The protocol’s O(1) filtering is robust regardless of the asset size or transaction complexity. The fixed 4-byte nLockTime acts as a high-entropy, past-timestamp beacon that allows indexers to scan billions of transactions efficiently.

By comparison, Ordinals and BRC-20 require parsing witness data for envelope patterns, while Counterparty/Omni rely on scanning OP\_RETURN outputs. Full indexer synchronization for these protocols can take 24–72 hours and tens of gigabytes of database space. Lockchain achieves the same discovery goals with minimal storage and computational overhead.

\section{Summary}
We describe the usage of the transaction header nLockTime field for O(1) protocol discovery by encoding metadata in the unused past-timestamp range. The Lockchain Protocol introduces a highly efficient discovery layer for Bitcoin assets. By repurposing the nLockTime field as a free 4-byte beacon, it enables the creation of secure, verifiable, and fractionally ownable on-chain data assets without the high indexing costs of other protocols.

The proofs-of-concept demonstrate systematic type routing via nLockTime value and transaction sequence numbering for multi-transaction protocols. If widely adopted, the protocols described here would lower the cost of launching and maintaining indexers for most existing meta-protocols.

The protocol assumes SegWit transactions and leverages the witness discount, though discovery depends only on the 4-byte header. Collisions in nLockTime values are harmless but can generate noise. A voluntary registry of Magic/Type allocations can further reduce collision noise and improve interoperability. The time-dependent nature of the usable nLockTime range is acknowledged as an edge condition. As time progresses, more past timestamps become available for signalling, expanding rather than shrinking the available space.

Repurposing nLockTime may be controversial in the context of anti-spam efforts on Bitcoin, but Lockchain shows that current protocol fields can be used efficiently for discovery without conflicting with consensus or transaction validity. Historical precedent supports field repurposing: nSequence, nVersion, and BIP implementations demonstrate that underutilized transaction fields can serve dual purposes, providing architectural efficiency without fundamental cryptographic innovation.

The reference implementation of the proposed protocol is available as open-source software on GitHub \url{https://github.com/softmatsg/btc-nlock-lighthouse}.

% --- Bibliography ---
\bibliographystyle{unsrt}
\bibliography{bibliography}

% --- Annexes ---
\newpage
\appendix
\section{Execution Logs: POC 1 (Parallel Sharding)}

\begin{lstlisting}
Protocol settings loaded.
Loaded 1120 bytes from 'bitcoin.txt'.
Executing POC 1 (Parallel Sharding)... Please wait.
======================================================================
POC1: PARALLEL SHARDING
Magic: 0x4C, Type: 0x01
Timestamp proof:
Timestamp: 1765721995
Hash: 6a2465281d638d3d49089914bcef6971935e538c1e2d2926da7e708c442d966e
Hash size: 32 bytes
File sharding:
File size: 1,120 bytes
OP_RETURN limit: 80 bytes
Shard payload: 48 bytes (limit - hash)
Shards: 24
Cost calculation (estimated):
Funding outputs: 23
Funding UTXO value: 1200 sats each (fee budget per shard)
Network fee per shard: 1200 sats
Total network fee: 28800 sats
Protocol fee: 2880 sats
Estimated funding TX fee: 3000 sats
Total needed (estimate): 33480 sats
Using provided address: tb1qe7fywud98jrsxj96cjqcmsczxxc5aaq73ppmqt
Verifying UTXOs exist on-chain...
[WARN] Stale UTXO: 53a6f5224b30a4c6...:5 (already spent)
[WARN] Error checking UTXO 53a6f5224b30a4c6...: Error when retrieving UTXO's
[WARN] Found 2 stale UTXOs in wallet cache
Verified usable: 7 UTXOs
UTXO filtering:
Available: 20 UTXOs
Usable (>1000 sats, can sign): 7 UTXOs
Funding TX fee calculation:
Inputs: 7
Outputs: 26
Calculated fee: 1621 sats
UTXO selection:
Selected: 7 UTXOs
Total: 132,356 sats
Outputs: 30,480 sats
Fee: 1,621 sats
Change: 100,255 sats
======================================================================
FUNDING TRANSACTION
Added 7 inputs
OP_RETURN: hash + shard 0 (80 bytes)
Added 23 funding outputs
Protocol fee: 2880 sats
Change: 100,255 sats
Funding TX summary:
Inputs: 132,356 sats
Outputs: 130,735 sats
Fee: 1,621 sats
Locktime: 0x4C010000
Output breakdown (26 outputs total):
Output 0: OP_RETURN (nulldata)
Output 1: Funding UTXO 1200 sats
...
Output 24: Protocol fee 2880 sats
Output 25: Change 100,255 sats
Signing with 7 keys...
Note: SegWit verification happens on broadcast
Broadcasting funding TX...
[OK] Funding TX: fff6f1aa3c18fdc6c97bb080098713ad3101cc46ecd82812963bab776d828747
View: https://mempool.space/testnet/tx/fff6f1aa3c18fdc6c97bb080098713ad3101cc46ecd82812963bab776d828747
[WAIT] Waiting 3 seconds for funding TX to propagate...
======================================================================
WAITING FOR FUNDING UTXOs
Waiting for 23 UTXO(s)...
TX: fff6f1aa3c18fdc6...
Address: tb1qe7fywud98jrsxj96...
Waiting for TX to appear in mempool...
TXID: fff6f1aa3c18fdc6...
[OK] Transaction in mempool!
Waiting for UTXOs to appear...
DEBUG: Transaction has 26 total outputs
DEBUG: Found 24 outputs to address tb1qe7fywud98jrsxj96...
DEBUG: Expected value: 1200 sats
DEBUG: Value distribution: {1200: 23, 100255: 1}
[OK] Found all 23 UTXO(s) after 1 attempts!
======================================================================
SHARD TRANSACTIONS
Building 23 shard TXs...
[OK] Shard 1: 3943798fa92c2e4f... (locktime: 0x4C010001)
[OK] Shard 2: 758d0594ff2feee5... (locktime: 0x4C010002)
...
[OK] Shard 23: 908fd526df0b76b0... (locktime: 0x4C010017)
======================================================================
POC1 COMPLETE
Funding TX: fff6f1aa3c18fdc6c97bb080098713ad3101cc46ecd82812963bab776d828747
Shards broadcast: 24/24
Key feature: Shard sequence in nLockTime byte 4
Shard 0: 0x4C010000
Shard 1: 0x4C010001
Shard N: 0x4C010017

-- SUCCESS ---
Funding TX: fff6f1aa3c18fdc6c97bb080098713ad3101cc46ecd82812963bab776d828747
Total shards: 24
Successful: 24
Network fee: 28800 sats
Protocol fee: 2880 sats
======================================================================
TEST 2: Verify nLockTime Sequence Numbers
Transaction: fff6f1aa3c18fdc6...
Locktime: 0x4C010000
Magic: 0x4C ('L')
Type: 0x01
Byte 3: 0x00
Sequence: 0x00 (shard 0)
...
[OK] All transactions share same magic (0x4C) and type (0x01)
[OK] Sequence numbers are complete: 0 through 23
======================================================================
TEST 3: Extract and Reassemble Data (OUT OF ORDER)
Simulating out-of-order transaction discovery:
Original order: [0, 1, 2, 3, 4, 5, 6, 7, 8, 9, 10, 11, 12, 13, 14, 15, 16, 17, 18, 19, 20, 21, 22, 23]
Shuffled order: [14, 11, 10, 21, 1, 12, 5, 9, 0, 6, 19, 15, 23, 16, 17, 4, 3, 13, 18, 7, 2, 22, 8, 20]
...
Reassembling shards in correct order...
Reassembled total: 1120 bytes
Original file size: 1120 bytes
[OK] Size matches
======================================================================
TEST 4: Verify Timestamp and Hash
Checking hash consistency across all 24 shards...
[OK] All shards contain identical hash
On-chain hash: 6a2465281d638d3d49089914bcef6971935e538c1e2d2926da7e708c442d966e
Verifying timestamp-included hash...
Timestamp from result: 1765721995
Original hash: 6a2465281d638d3d49089914bcef6971935e538c1e2d2926da7e708c442d966e
On-chain hash: 6a2465281d638d3d49089914bcef6971935e538c1e2d2926da7e708c442d966e
Reconstructing hash from reassembled data + timestamp...
Reconstructed: 6a2465281d638d3d49089914bcef6971935e538c1e2d2926da7e708c442d966e
On-chain:      6a2465281d638d3d49089914bcef6971935e538c1e2d2926da7e708c442d966e
Match: True
[OK] HASH VERIFIED
The on-chain hash correctly includes the timestamp
======================================================================
TEST 5: Data Integrity Check
Comparing reassembled data with original file...
[OK] PERFECT MATCH
Reassembled data is byte-for-byte identical to original
======================================================================
VERIFICATION SUMMARY
POC1 demonstrates:
[OK] nLockTime sequence numbering (0x00 through 0x17)
[OK] Protocol discovery via magic byte (0x4C = 'L')
[OK] Transaction type identification (0x01 = POC1)
[OK] Out-of-order shard reassembly using sequence numbers
[OK] Timestamp-included hash for proof of existence
[OK] Data integrity verification across all shards
\end{lstlisting}

\newpage
\section{Execution Logs: POC 2 (Single-Tx Static Asset)}

\begin{lstlisting}
Protocol settings loaded.
Loaded 184,292 bytes from 'bitcoin.pdf'.
======================================================================
POC2: SINGLE-TX STATIC ASSET
Configuration:
File size: 184,292 bytes
Requested chunks: 10
Calculated chunk size: 18,430 bytes per chunk
Protocol headers:
Magic: 0x4C
Type: 0x02
Variant: 0x73 ('s')
Sequence: 0x01
Locktime: 0x4C027301
File processing:
Expected chunks: 10
Actual chunks: 10
Last chunk size: 18422 bytes
Merkle tree:
Hash method: sha256
Merkle root: 8d75277f6f4a80338fd7046eb83a39dee6a5fcfc3ee4dd7449a05d22c48e1218
Root size: 32 bytes
[OK] Using calculated fee: 2000 sats
Cost calculation:
Dust tokens: 10 x 650 = 6,500 sats
Protocol fee: 650 sats (10% of dust, min 650)
Total outputs: 7,150 sats
Network fee: 2,000 sats
Total needed: 9,150 sats
Estimated needed: 9,150 sats
Using provided address: tb1q3ln06dkdq5y43y8jvw0lq87nt5gsfhswvsm2u2
Verifying UTXOs exist on-chain...
[WARN] Stale UTXO: ebce419e9c7b93c7...:12 (already spent)
[WARN] Stale UTXO: ebce419e9c7b93c7...:25 (already spent)
[WARN] Found 2 stale UTXOs in wallet cache
Verified usable: 9 UTXOs
UTXO filtering:
Available: 31 UTXOs
Usable (>1000 sats, can sign): 9 UTXOs
Recalculated fee based on 4 inputs:
Estimated fee: 798 sats
Final fee (after testnet rules): 1200 sats
Final TX breakdown:
Selected: 4 UTXOs
Total input: 12,000 sats
Total output: 7,150 sats
Network fee: 1200 sats
Change: 3,650 sats
======================================================================
BUILDING GENESIS TRANSACTION
Added 4 inputs
Output 0: OP_RETURN with 32-byte Merkle root
Outputs 1-10: 10 ownership tokens @ 650 sats each
Output 11: Protocol fee 650 sats
Output 12: Change 3,650 sats
Transaction totals:
Input total: 12,000 sats
Output total: 10,800 sats
Fee: 1,200 sats
Verifying dust token outputs:
Token 0: 650 sats, script_type=p2wpkh, address=tb1q3ln06dkdq5y43y8j...
...
Token 9: 650 sats, script_type=p2wpkh, address=tb1q3ln06dkdq5y43y8j...
Signing transaction...
Signing with 4 keys...
Note: SegWit verification happens on broadcast
Transaction size: 1020 bytes
Broadcasting transaction...
[OK] Genesis TX: 3501499ea2c5db806679a60d2683a6dffa244499407947233df25684e9bcba7e
View: https://mempool.space/testnet/tx/3501499ea2c5db806679a60d2683a6dffa244499407947233df25684e9bcba7e
======================================================================
[OK] SUCCESS - STATIC ASSET CREATED
Asset details:
Genesis TXID: 3501499ea2c5db806679a60d2683a6dffa244499407947233df25684e9bcba7e
Asset ID: 3501499ea2c5db806679a60d2683a6dffa244499407947233df25684e9bcba7e
File size: 184,292 bytes
Chunks: 10
Chunk size: 18,430 bytes
Ownership tokens: 10 @ 650 sats each
Merkle root: 8d75277f6f4a80338fd7046eb83a39dee6a5fcfc3ee4dd7449a05d22c48e1218
Locktime structure:
[0x4C][0x02][0x73][0x01]
['L'][Type]['s'][Seq 1]
Ownership:
All 10 tokens issued to: tb1q3ln06dkdq5y43y8jvw0lq87nt5gsfhswvsm2u2
Token indices: Outputs 1-10 in genesis TX
Costs:
Network fee: 1,200 sats
Protocol fee: 650 sats
Total cost: 1,850 sats
Key insight:
Cost based on num_chunks (10), NOT file size!
Same cost for 1KB or 1GB file with 10 chunks
======================================================================
[OK] COMPLETE
TXID: 3501499ea2c5db806679a60d2683a6dffa244499407947233df25684e9bcba7e
View: https://mempool.space/testnet/tx/3501499ea2c5db806679a60d2683a6dffa244499407947233df25684e9bcba7e
====
VERIFYING DUST TOKEN LOCKING SCRIPTS
Genesis TXID: 03af0a9882dd5bb151ab7db4438b183c448d47963e03bb8d8577760a305c4835
Expected tokens: 10
Issuance address: tb1q3ln06dkdq5y43y8jvw0lq87nt5gsfhswvsm2u2
Fetching transaction from blockchain...
[OK] Transaction found
Transaction has 13 outputs:
Output 0 (Merkle root):
Script type: nulldata
[OK] Correct (OP_RETURN)
Dust tokens (Outputs 1-10):
Token 0 (Output 1):
Value: 546 sats
Script type: p2wpkh
Address: tb1q3ln06dkdq5y43y8jvw0lq87nt5...
[OK] Value correct (546 sats)
[OK] Script type correct (P2WPKH SegWit)
[OK] Address correct
[OK] Spendable (wallet has private key)
...
======================================================================
[OK] ALL DUST TOKENS VERIFIED SUCCESSFULLY
All 10 tokens are:
 - Standard P2WPKH (SegWit native)
 - Exactly 546 sats (dust limit)
 - Locked to your address: tb1q3ln06dkdq5y43y8jvw0lq87nt5...
 - Spendable by your wallet
These tokens can be transferred like regular Bitcoin UTXOs!
Token outputs stored in 'token_outputs' variable.
Example: token_outputs[0] = first token
\end{lstlisting}

\newpage
\section{Execution Logs: POC 2.5 (Two-Transaction Static Asset)}

\begin{lstlisting}
Protocol settings loaded.
Loaded 184,292 bytes from 'bitcoin.pdf'.
======================================================================
POC2.5: TWO-TX STATIC ASSET WITH BINDING
Configuration:
File size: 184,292 bytes
Requested chunks: 10
Calculated chunk size: 18,430 bytes per chunk
Protocol headers:
Magic: 0x4C
Type: 0x03
Variant: 0x74 ('t')
Genesis sequence: 0x00
Tokenization sequence: 0x01
File processing:
Expected chunks: 10
Actual chunks: 10
Merkle tree:
Merkle root: 8d75277f6f4a80338fd7046eb83a39dee6a5fcfc3ee4dd7449a05d22c48e1218
======================================================================
COST CALCULATION
Protocol fees (ALL collected in TX1):
TX1 component: 200 sats (10% of DEFAULT_FEE)
TX2 component: 650 sats (10% of dust)
Total protocol fee: 850 sats
(Minimum: 650 sats to avoid dust output)
TX1 (Genesis):
Network fee: 2,000 sats
Protocol fee: 850 sats (collected here)
Total: 2,850 sats
TX2 (Tokenization):
Dust outputs: 10 x 650 = 6,500 sats
Protocol fee: 0 sats (collected in TX1)
Network fee: 2,000 sats
Total: 8,500 sats
Overall cost:
Total protocol fees: 850 sats
Total network fees: 4000 sats
Grand total (estimate): 11,350 sats
Using provided address: tb1q3ln06dkdq5y43y8jvw0lq87nt5gsfhswvsm2u2
======================================================================
TX1: GENESIS (Merkle Root Commitment)
Verifying UTXOs exist on-chain...
[WARN] Stale UTXO: ebce419e9c7b93c7...:1 (already spent)
...
[WARN] Found 5 stale UTXOs in wallet cache
Verified usable: 6 UTXOs
UTXO filtering:
Available: 31 UTXOs
Usable (>1000 sats, can sign): 6 UTXOs
TX1 UTXO selection:
Selected: 1 UTXOs
Total: 3,000 sats
Change: 150 sats
Added 1 inputs
Output 0: OP_RETURN with Merkle root
Output 1: Protocol fee 850 sats (TX1+TX2 combined)
[WARN] Warning: Change 150 sats below dust, will become fee
TX1 totals:
Inputs: 3,000 sats
Outputs: 850 sats
Fee: 2,150 sats
Locktime: 0x4C037400
Signing TX1 with 1 keys...
Note: SegWit verification happens on broadcast
Broadcasting TX1...
[OK] Genesis TX: 7d6b5dc9165e43ca7202b43d3d1499e173b2720b8efdf693835a66f3d4fff535
View: https://mempool.space/testnet/tx/7d6b5dc9165e43ca7202b43d3d1499e173b2720b8efdf693835a66f3d4fff535
[WAIT] Waiting 3 seconds for TX1 to propagate...
======================================================================
CRYPTOGRAPHIC BINDING
Binding hash = SHA256(genesis_txid + merkle_root)
Genesis TXID: 7d6b5dc9165e43ca7202b43d3d1499e173b2720b8efdf693835a66f3d4fff535
Merkle root: 8d75277f6f4a80338fd7046eb83a39dee6a5fcfc3ee4dd7449a05d22c48e1218
Binding hash: edd84e2a7ff067d5735d8e923c984f882b3c136e718bc83bac1fdd12c7aa75c8
[OK] Binding verified
======================================================================
TX2: TOKENIZATION (Dust Outputs + Binding)
Refreshing wallet for TX1 change...
[WARN] WARNING: Full wallet rescan requested
This may trigger rate limiting!
TX2 UTXO selection:
Selected: 3 UTXOs
Total: 9,000 sats
Change: 500 sats
Added 3 inputs
Output 0: OP_RETURN with binding hash
Outputs 1-10: 10 ownership tokens @ 650 sats each
TX2 totals:
Inputs: 9,000 sats
Outputs: 6,500 sats
Fee: 2,500 sats
Locktime: 0x4C037401
Signing TX2 with 3 keys...
Note: SegWit verification happens on broadcast
Broadcasting TX2...
[OK] Tokenization TX: 8b7e854df00b579610ccb43e5eb390c2a99e35451bbd9681bae9e7187c8b38ad
View: https://mempool.space/testnet/tx/8b7e854df00b579610ccb43e5eb390c2a99e35451bbd9681bae9e7187c8b38ad
======================================================================
[OK] SUCCESS - TWO-TX ASSET CREATED
Asset details:
Genesis TXID: 7d6b5dc9165e43ca7202b43d3d1499e173b2720b8efdf693835a66f3d4fff535
Tokenization TXID: 8b7e854df00b579610ccb43e5eb390c2a99e35451bbd9681bae9e7187c8b38ad
Asset ID: 7d6b5dc9165e43ca7202b43d3d1499e173b2720b8efdf693835a66f3d4fff535
File size: 184,292 bytes
Chunks: 10
Chunk size: 18,430 bytes
Ownership tokens: 10 @ 650 sats each
Cryptographic binding:
Merkle root: 8d75277f6f4a80338fd7046eb83a39dee6a5fcfc3ee4dd7449a05d22c48e1218
Binding hash: b'\xed\xd8N*\x7f\xf0g\xd5s]\x8e\x92<\x98O\x88+<\x13nq\x8b\xc8;\xac\x1f\xdd\x12\xc7\xaau\xc8'
Formula: SHA256(genesis_txid + merkle_root)
Locktime sequences:
TX1 (Genesis): 0x4C037400 (sequence 0)
TX2 (Tokenization): 0x4C037401 (sequence 1)
Costs:
TX1 fee: 2,150 sats
TX2 fee: 2,500 sats
Protocol fees: 850 sats
Total cost: 5,500 sats
======================================================================
[OK] COMPLETE
Genesis TXID: 7d6b5dc9165e43ca7202b43d3d1499e173b2720b8efdf693835a66f3d4fff535
Tokenization TXID: 8b7e854df00b579610ccb43e5eb390c2a99e35451bbd9681bae9e7187c8b38ad
Asset ID: 7d6b5dc9165e43ca7202b43d3d1499e173b2720b8efdf693835a66f3d4fff535
======================================================================
VERIFYING POC2.5 BINDING AND SEQUENCES
...
[OK] Both transactions found
======================================================================
TEST 1: Cryptographic Binding
Binding formula: SHA256(genesis_txid + merkle_root)
Genesis TXID: 7d6b5dc9165e43ca7202b43d3d1499e173b2720b8efdf693835a66f3d4fff535
Merkle root: 8d75277f6f4a80338fd7046eb83a39dee6a5fcfc3ee4dd7449a05d22c48e1218
Expected binding hash: edd84e2a7ff067d5735d8e923c984f882b3c136e718bc83bac1fdd12c7aa75c8
[OK] BINDING VERIFIED
The binding hash correctly links the issuer (TX1) to the content (Merkle root)
Verifying binding hash in TX2...
On-chain binding hash: edd84e2a7ff067d5735d8e923c984f882b3c136e718bc83bac1fdd12c7aa75c8
Expected binding hash: edd84e2a7ff067d5735d8e923c984f882b3c136e718bc83bac1fdd12c7aa75c8
[OK] Binding hash matches on-chain data
======================================================================
TEST 2: nLockTime Sequence Verification
TX1 (Genesis):
Locktime: 0x4C037400
Magic: 0x4C
Type: 0x03
Byte 3: 0x74
Sequence: 0x00
[OK] TX1 sequence correct (0x00 for Genesis)
TX2 (Tokenization):
Locktime: 0x4C037401
Magic: 0x4C
Type: 0x03
Byte 3: 0x74
Sequence: 0x01
[OK] TX2 sequence correct (0x01 for Tokenization)
[OK] Sequences are sequential (TX1=0x00, TX2=0x01)
======================================================================
TEST 3: Merkle Root in TX1
On-chain Merkle root: 8d75277f6f4a80338fd7046eb83a39dee6a5fcfc3ee4dd7449a05d22c48e1218
Expected Merkle root: 8d75277f6f4a80338fd7046eb83a39dee6a5fcfc3ee4dd7449a05d22c48e1218
[OK] Merkle root matches on-chain data
======================================================================
TEST 4: Dust Tokens in TX2
Expected 10 dust tokens (outputs 1-10):
Token 0: [OK] [OK] [OK] (value=650, type=p2wpkh)
Token 1: [OK] [OK] [OK] (value=650, type=p2wpkh)
...
[OK] All 10 tokens are valid
======================================================================
VERIFICATION SUMMARY
POC2.5 demonstrates:
[OK] Two-transaction model (Genesis + Tokenization)
[OK] Cryptographic binding (SHA256(genesis_txid + merkle_root))
[OK] Sequence tracking in nLockTime (TX1=0x00, TX2=0x01)
[OK] Merkle root commitment in TX1
[OK] Binding hash in TX2 OP_RETURN
[OK] Standard P2WPKH dust tokens
The binding ensures that:
TX2 tokens are cryptographically linked to TX1 issuer
Cannot fake issuer identity (genesis_txid is unique)
Cannot swap content (merkle_root is in binding)
Indexers can verify the complete chain
\end{lstlisting}

\newpage
\section{Execution Logs: POC 3 (Multisig-Protected Assets)}

\begin{lstlisting}
Loaded existing wallet: poc3_wallet_002
Lockchain Client initialized
Wallet: poc3_wallet_002
Network: testnet
Main address: tb1q7ymj8cht3tw5ypavxs9y0e35yn06ap0zt5y58x
======================================================================
POC3: MULTISIG-PROTECTED ASSETS
Loading SAVED POC3 keys...
[OK] Issuer:    tb1q4ex6vcwkhg3vm9zhhyq0ftfqu3y52heyxku38j
[OK] Service:   tb1q7ymj8cht3tw5ypavxs9y0e35yn06ap0zt5y58x
[OK] Recipient: tb1qd2nev6w7syjwz02fdp3rnv9n0evd39wkmxalcf
[OK] Keys loaded successfully from wallet
======================================================================
Ready for POC3 testing!
======================================================================
CHECKING WALLET BALANCE
Wallet addresses:
Main wallet:  tb1q7ymj8cht3tw5ypavxs9y0e35yn06ap0zt5y58x
Issuer:       tb1q4ex6vcwkhg3vm9zhhyq0ftfqu3y52heyxku38j
Service:      tb1q7ymj8cht3tw5ypavxs9y0e35yn06ap0zt5y58x
Recipient:    tb1qd2nev6w7syjwz02fdp3rnv9n0evd39wkmxalcf
Cached wallet balance: 483,273.0 sats
[OK] Sufficient balance for testing
Checking UTXO distribution...
tb1q0pksfq2kkxa69tj2t7hh74gekgge6jkdauf6ge : 189,198 sats
Issuer:       294,075 sats
======================================================================
Creating multisig-protected asset...
File: bitcoin.txt
Size: 184,292 bytes
Chunks: 10
======================================================================
POC3: MULTISIG-PROTECTED ASSET
Magic: 0x4C, Type: 0x03
Variants: Genesis='g'(0x67), Tokenization='t'(0x74), Transfer='x'(0x78)
File chunking (by num_chunks):
File size: 184,292 bytes
Requested chunks: 10
Chunk size: 18,430 bytes
Merkle tree:
Hash method: sha256 (default)
Merkle root: 8d75277f6f4a80338fd7046eb83a39dee6a5fcfc3ee4dd7449a05d22c48e1218
Root size: 32 bytes
[WARN] TESTNET: TX1 outputs (650 sats) <= dust limit
TX1 fee set to 0 (dust rule)
Cost calculation:
TX1 (Genesis):
Protocol fee: 650 sats
Network fee:  0 sats
Total:        650 sats
TX2 (Tokenization):
Tokens value: 6,500 sats (10 x 650)
Network fee:  2000 sats
Total:        8,500 sats
Grand total: 9,150 sats
Using provided issuance address: tb1q4ex6vcwkhg3vm9zhhyq0ftfqu3y52heyxku38j
Using provided service pubkey: 038ca054840e4bb0124b9bb7569e4653...
Issuer pubkey: 0229b891c842e92514cd8782b5c03cd4...
Updating wallet state from blockchain...
[OK] Wallet updated
======================================================================
TX1: GENESIS (Merkle Root Commitment)
TX1 Locktime: 0x4C036700
[4C][03][67(g)][00]
Selecting UTXOs for TX1...
Cached UTXOs: 3
[OK] Usable UTXO: 0b47c9564f337ffd08e921c4b945c7905d180bfa98c87c29d4af1f7959006cbf:0 - 189,198 sats (389 confs) at tb1q0pksfq2kkxa69tj2t7hh74gekgge6jkdauf6ge
[OK] Usable UTXO: 3838050feab21f78e0af435cb48b0d2c24a1bdacde47733cbb0d63ad56bb1dde:10 - 114,246 sats (7 confs) at tb1q4ex6vcwkhg3vm9zhhyq0ftfqu3y52heyxku38j
[OK] Usable UTXO: b7b7add2050572f018eda33f8f7304d8994a8d3783f6557fd79bc0c5512bfe92:10 - 179,829 sats (377 confs) at tb1q4ex6vcwkhg3vm9zhhyq0ftfqu3y52heyxku38j
Found 3 usable UTXOs
Verifying cached UTXOs exist on-chain...
TX1 UTXO selection:
Selected: 1 UTXOs
Total value: 114,246 sats
Needed: 650 sats
Fee recalculation:
Estimated fee: 0 sats
Recalculated (1 inputs, 3 outputs): 219 sats
Final fee: 219 sats
Change: 113,377 sats
Building Genesis TX...
TX1 structure:
Inputs: 1
Outputs: 3
0: OP_RETURN (Merkle root)
1: Protocol fee (650 sats)
2: Change (113377 sats)
Signing Genesis TX...
Broadcasting Genesis TX...
[OK] Genesis TX broadcast successful!
TXID: e78c935ace261d1f987c451264e4934982d13eec9f02f9c7db918326cec34176
View: https://mempool.space/testnet/tx/e78c935ace261d1f987c451264e4934982d13eec9f02f9c7db918326cec34176
Binding hash: 72a83572bbe654e6c4fa836527e2c4287f25c3308961184b3178e6108beca936
= SHA256(genesis_txid + merkle_root)
======================================================================
TX2: TOKENIZATION (Multisig-Protected Tokens)
TX2 Locktime: 0x4C037401
[4C][03][74(t)][01]
Waiting for TX1 output UTXO...
Expecting: e78c935ace261d1f987c451264e4934982d13eec9f02f9c7db918326cec34176:2
Address: tb1q4ex6vcwkhg3vm9zhhyq0ftfqu3y52heyxku38j
Value: 113377 sats
Waiting for 1 UTXO(s)...
TX: e78c935ace261d1f...
Address: tb1q4ex6vcwkhg3vm9zh...
Waiting for TX to appear in mempool...
TXID: e78c935ace261d1f...
[OK] Transaction in mempool!...
Waiting for UTXOs to appear...
DEBUG: Transaction has 3 total outputs
DEBUG: Found 1 outputs to address tb1q4ex6vcwkhg3vm9zh...
DEBUG: Expected value: 113377 sats
DEBUG: Value distribution: {113377: 1}
[OK] Found all 1 UTXO(s) after 1 attempts!
[OK] Found TX1 output: 113,377 sats
Selecting UTXOs for TX2...
TX2 change: 104,877 sats
Building Tokenization TX...
Creating 10 multisig-protected tokens...
TX2 structure:
Inputs: 1
Outputs: 11
0: Token 0 (650 sats, 2-of-2 multisig)
...
10: Change (104877 sats)
Signing Tokenization TX...
Broadcasting Tokenization TX...
[OK] Tokenization TX broadcast successful!
TXID: d61df2674ca590fbdaa7b50168dbb151fdc4d52960f692be6e2ddbef0b527db1
View: https://mempool.space/testnet/tx/d61df2674ca590fbdaa7b50168dbb151fdc4d52960f692be6e2ddbef0b527db1
Waiting for token outputs to be visible...
(This ensures transfers can find the UTXOs)
Waiting for 1 UTXO(s)...
TX: d61df2674ca590fb...
Address: tb1q3fs3pr03mx4xhwna...
Waiting for TX to appear in mempool...
TXID: d61df2674ca590fb...
[OK] Transaction in mempool!
Waiting for UTXOs to appear...
DEBUG: Transaction has 11 total outputs
DEBUG: Found 10 outputs to address tb1q3fs3pr03mx4xhwna...
DEBUG: Expected value: 650 sats
DEBUG: Value distribution: {650: 10}
[OK] Found all 1 UTXO(s) after 1 attempts!
[OK] Token outputs are visible in mempool!
Waiting for TX2 change output (potential fee source)...
Waiting for 1 UTXO(s)...
TX: d61df2674ca590fb...
Address: tb1q4ex6vcwkhg3vm9zh...
Waiting for TX to appear in mempool...
TXID: d61df2674ca590fb...
[OK] Transaction in mempool!
Waiting for UTXOs to appear...
DEBUG: Transaction has 11 total outputs
DEBUG: Found 1 outputs to address tb1q4ex6vcwkhg3vm9zh...
DEBUG: Expected value: 104877 sats
DEBUG: Value distribution: {104877: 1}
[OK] Found all 1 UTXO(s) after 1 attempts!
[OK] TX2 change output visible! (104877 sats)
You can use this as 'fee_source_utxo' for the transfer.
Updating wallet state to mark spent inputs...
[OK] MULTISIG-PROTECTED ASSET CREATED
Asset details:
Genesis TXID:       e78c935ace261d1f987c451264e4934982d13eec9f02f9c7db918326cec34176
Tokenization TXID: d61df2674ca590fbdaa7b50168dbb151fdc4d52960f692be6e2ddbef0b527db1
Asset ID:           e78c935ace261d1f987c451264e4934982d13eec9f02f9c7db918326cec34176
File size:          184,292 bytes
Chunks:             10
Chunk size:         18,430 bytes
Merkle root:        8d75277f6f4a80338fd7046eb83a39dee6a5fcfc3ee4dd7449a05d22c48e1218
Binding hash:       72a83572bbe654e6c4fa836527e2c4287f25c3308961184b3178e6108beca936
Locktime structure:
Genesis:       0x4C036700 - variant 'g' (genesis)
Tokenization: 0x4C037401 - variant 't' (tokenization)
Transfers:    0x4C0378XX - variant 'x' (transfer), XX = transfer count
Examples:
1st transfer: 0x4C037801 (sequence = 0x01)
2nd transfer: 0x4C037802 (sequence = 0x02)
3rd transfer: 0x4C037803 (sequence = 0x03)
etc...
Token protection:
Type: 2-of-2 multisig (Issuer + Service)
All 10 tokens require service co-signature
Prevents accidental spending by standard wallets
Costs:
TX1 fee:      219 sats
TX2 fee:      2,000 sats
Protocol fee: 650 sats
Total cost:   2,869 sats
Key insight:
Tokens are SAFE from accidental spending!
Service validates transfers before co-signing
======================================================================
ASSET CREATED SUCCESSFULLY
Genesis TX:       e78c935ace261d1f987c451264e4934982d13eec9f02f9c7db918326cec34176
Tokenization TX: d61df2674ca590fbdaa7b50168dbb151fdc4d52960f692be6e2ddbef0b527db1
Asset ID:         e78c935ace261d1f987c451264e4934982d13eec9f02f9c7db918326cec34176
...
======================================================================
TOKEN 0 DETAILS
Index:           0
UTXO:            d61df2674ca590fbdaa7b50168dbb151fdc4d52960f692be6e2ddbef0b527db1:0
Value:           650 sats
Address:         tb1q3fs3pr03mx4xhwna3luj5dqltjem0tlt72mjnsklxweucvc3fmwq9xuskx
Protection:
Type:            2-of-2 multisig
Owner Pubkey:    0229b891c842e92514cd8782b5c03cd48eb01703d0fd1c2a9e36577e4b70793a3b
Service Pubkey: 038ca054840e4bb0124b9bb7569e4653d35aeb74c01ee1a5631a76e947fb904eb7
Transfer Count: 0
Redeem Script:
52210229b891c842e92514cd8782b5c03cd48eb01703d0fd1c2a9e36577e4b70793a3b21038ca054840e4bb0124b9bb7569e4653d35aeb74c01ee1a5631a76e947fb904eb752ae
View on blockchain:https://mempool.space/testnet/tx/d61df2674ca590fbdaa7b50168dbb151fdc4d52960f692be6e2ddbef0b527db1
Key insight:
This token CANNOT be spent by a standard wallet!
It requires BOTH issuer AND service signatures.
======================================================================
PREPARING FOR TRANSFER
Current asset to transfer from:
Genesis TX: e78c935ace261d1f987c451264e4934982d13eec9f02f9c7db918326cec34176
Token TX:   d61df2674ca590fbdaa7b50168dbb151fdc4d52960f692be6e2ddbef0b527db1
Tokens:     10
Querying blockchain for available UTXOs...
Checking address: tb1q4ex6vcwkhg3vm9zhhyq0ftfqu3y52heyxku38j
Found 2 UTXO(s) at issuer address:
UTXO: b7b7add2050572f018eda33f8f7304d8994a8d3783f6557fd79bc0c5512bfe92:10
Value: 179,829 sats
[OK] Suitable for fee source!
======================================================================
[OK] FEE SOURCE READY
TXID: b7b7add2050572f018eda33f8f7304d8994a8d3783f6557fd79bc0c5512bfe92
Vout: 10
Value: 179,829 sats
Address: tb1q4ex6vcwkhg3vm9zhhyq0ftfqu3y52heyxku38j
======================================================================
TRANSFERRING TOKEN 0
Transfer details:
From: tb1q4ex6vcwkhg3vm9zhhyq0ftfqu3y52heyxku38j
To:   tb1qd2nev6w7syjwz02fdp3rnv9n0evd39wkmxalcf
Token: #0
======================================================================
POC3: TOKEN TRANSFER
Asset ID: e78c935ace261d1f987c451264e4934982d13eec9f02f9c7db918326cec34176
Token UTXO: d61df2674ca590fbdaa7b50168dbb151fdc4d52960f692be6e2ddbef0b527db1:0
Current transfer count: 0
Verifying token UTXO exists and is spendable...
[OK] Token TX found!
TX: d61df2674ca590fb...
Output: 0
Value: 650 sats
Address: tb1q3fs3pr03mx4xhwna3luj5dqltj...
Checking if output is unspent (including mempool)...
[WARN] Token UTXO is UNCONFIRMED (0 confirmations)
This is normal for newly created tokens.
The UTXO exists in mempool and will be spendable soon.
Transfer TX Locktime: 0x4C037801
[4C][03][78(x)][01]
Variant: 'x' (transfer)
Transfer count: 0 -> 1
New token owner: 0213060c351e7f4c845c64768c1665d1...
New token address: tb1qqwx9nv56lk9an8he7qa24jphsveydyt4hynmrn6r9paqp4t9phqsgg6vc2
Using fee source UTXO: d61df2674ca590fbdaa7b50168dbb151fdc4d52960f692be6e2ddbef0b527db1:10
Performing deep check on fee source...
Verifying fee source UTXO is unspent...
[OK] Fee source UTXO is unspent
Change: 104540 sats
Transfer TX structure:
Inputs: 2
0: Token UTXO (650 sats, 2-of-2 multisig)
1: Fee source (104877 sats)
Outputs: 2
0: New token (650 sats, 2-of-2 multisig)
1: Change (104540 sats)
Fee: 337 sats
!!!!!!!!!!!!!!!!!!!!!!!!!!!!!!!!!!!!!!!!!!!!!!!!!!!!!!!!!!!!
DEBUG: PRE-BROADCAST CHECKLIST
Check these UTXOs on https://mempool.space/testnet
INPUT 0 (Token):
TXID: d61df2674ca590fbdaa7b50168dbb151fdc4d52960f692be6e2ddbef0b527db1
VOUT: 0
INPUT 1 (Fee Source):
TXID: d61df2674ca590fbdaa7b50168dbb151fdc4d52960f692be6e2ddbef
\end{lstlisting}

\end{document}